\documentclass[aps,prl,preprint,tightenlines,superscriptaddress,showpacs,byrevtex]{revtex4}

\usepackage{notes2bib}
\usepackage{graphicx}
\usepackage{gensymb}
\usepackage{url}
\usepackage{graphicx}
\usepackage{amsmath}
\usepackage{latexsym}
\usepackage{amssymb}
\usepackage{color}

\usepackage{slashed}
\newcommand{\cost}{ {\rm cos} \theta}

\begin{document}

\preprint{\vbox{ \hbox{   }
                 \hbox{BELLE-CONF-1611}
}}


\title{\quad\\[0.5cm]Observation of Transverse $\Lambda/\bar{\Lambda}$ Hyperon Polarization in $e^+e^-$ Annihilation at Belle}
\noaffiliation
\affiliation{Aligarh Muslim University, Aligarh 202002}
\affiliation{University of the Basque Country UPV/EHU, 48080 Bilbao}
\affiliation{Beihang University, Beijing 100191}
\affiliation{University of Bonn, 53115 Bonn}
\affiliation{Budker Institute of Nuclear Physics SB RAS, Novosibirsk 630090}
\affiliation{Faculty of Mathematics and Physics, Charles University, 121 16 Prague}
\affiliation{Chiba University, Chiba 263-8522}
\affiliation{Chonnam National University, Kwangju 660-701}
\affiliation{University of Cincinnati, Cincinnati, Ohio 45221}
\affiliation{Deutsches Elektronen--Synchrotron, 22607 Hamburg}
\affiliation{University of Florida, Gainesville, Florida 32611}
\affiliation{Department of Physics, Fu Jen Catholic University, Taipei 24205}
\affiliation{Justus-Liebig-Universit\"at Gie\ss{}en, 35392 Gie\ss{}en}
\affiliation{Gifu University, Gifu 501-1193}
\affiliation{II. Physikalisches Institut, Georg-August-Universit\"at G\"ottingen, 37073 G\"ottingen}
\affiliation{SOKENDAI (The Graduate University for Advanced Studies), Hayama 240-0193}
\affiliation{Gyeongsang National University, Chinju 660-701}
\affiliation{Hanyang University, Seoul 133-791}
\affiliation{University of Hawaii, Honolulu, Hawaii 96822}
\affiliation{High Energy Accelerator Research Organization (KEK), Tsukuba 305-0801}
\affiliation{J-PARC Branch, KEK Theory Center, High Energy Accelerator Research Organization (KEK), Tsukuba 305-0801}
\affiliation{Hiroshima Institute of Technology, Hiroshima 731-5193}
\affiliation{IKERBASQUE, Basque Foundation for Science, 48013 Bilbao}
\affiliation{University of Illinois at Urbana-Champaign, Urbana, Illinois 61801}
\affiliation{Indian Institute of Science Education and Research Mohali, SAS Nagar, 140306}
\affiliation{Indian Institute of Technology Bhubaneswar, Satya Nagar 751007}
\affiliation{Indian Institute of Technology Guwahati, Assam 781039}
\affiliation{Indian Institute of Technology Madras, Chennai 600036}
\affiliation{Indiana University, Bloomington, Indiana 47408}
\affiliation{Institute of High Energy Physics, Chinese Academy of Sciences, Beijing 100049}
\affiliation{Institute of High Energy Physics, Vienna 1050}
\affiliation{Institute for High Energy Physics, Protvino 142281}
\affiliation{Institute of Mathematical Sciences, Chennai 600113}
\affiliation{INFN - Sezione di Torino, 10125 Torino}
\affiliation{Advanced Science Research Center, Japan Atomic Energy Agency, Naka 319-1195}
\affiliation{J. Stefan Institute, 1000 Ljubljana}
\affiliation{Kanagawa University, Yokohama 221-8686}
\affiliation{Institut f\"ur Experimentelle Kernphysik, Karlsruher Institut f\"ur Technologie, 76131 Karlsruhe}
\affiliation{Kavli Institute for the Physics and Mathematics of the Universe (WPI), University of Tokyo, Kashiwa 277-8583}
\affiliation{Kennesaw State University, Kennesaw, Georgia 30144}
\affiliation{King Abdulaziz City for Science and Technology, Riyadh 11442}
\affiliation{Department of Physics, Faculty of Science, King Abdulaziz University, Jeddah 21589}
\affiliation{Korea Institute of Science and Technology Information, Daejeon 305-806}
\affiliation{Korea University, Seoul 136-713}
\affiliation{Kyoto University, Kyoto 606-8502}
\affiliation{Kyungpook National University, Daegu 702-701}
\affiliation{\'Ecole Polytechnique F\'ed\'erale de Lausanne (EPFL), Lausanne 1015}
\affiliation{P.N. Lebedev Physical Institute of the Russian Academy of Sciences, Moscow 119991}
\affiliation{Faculty of Mathematics and Physics, University of Ljubljana, 1000 Ljubljana}
\affiliation{Ludwig Maximilians University, 80539 Munich}
\affiliation{Luther College, Decorah, Iowa 52101}
\affiliation{University of Maribor, 2000 Maribor}
\affiliation{Max-Planck-Institut f\"ur Physik, 80805 M\"unchen}
\affiliation{School of Physics, University of Melbourne, Victoria 3010}
\affiliation{Middle East Technical University, 06531 Ankara}
\affiliation{University of Miyazaki, Miyazaki 889-2192}
\affiliation{Moscow Physical Engineering Institute, Moscow 115409}
\affiliation{Moscow Institute of Physics and Technology, Moscow Region 141700}
\affiliation{Graduate School of Science, Nagoya University, Nagoya 464-8602}
\affiliation{Kobayashi-Maskawa Institute, Nagoya University, Nagoya 464-8602}
\affiliation{Nara University of Education, Nara 630-8528}
\affiliation{Nara Women's University, Nara 630-8506}
\affiliation{National Central University, Chung-li 32054}
\affiliation{National United University, Miao Li 36003}
\affiliation{Department of Physics, National Taiwan University, Taipei 10617}
\affiliation{H. Niewodniczanski Institute of Nuclear Physics, Krakow 31-342}
\affiliation{Nippon Dental University, Niigata 951-8580}
\affiliation{Niigata University, Niigata 950-2181}
\affiliation{University of Nova Gorica, 5000 Nova Gorica}
\affiliation{Novosibirsk State University, Novosibirsk 630090}
\affiliation{Osaka City University, Osaka 558-8585}
\affiliation{Osaka University, Osaka 565-0871}
\affiliation{Pacific Northwest National Laboratory, Richland, Washington 99352}
\affiliation{Panjab University, Chandigarh 160014}
\affiliation{Peking University, Beijing 100871}
\affiliation{University of Pittsburgh, Pittsburgh, Pennsylvania 15260}
\affiliation{Punjab Agricultural University, Ludhiana 141004}
\affiliation{Research Center for Electron Photon Science, Tohoku University, Sendai 980-8578}
\affiliation{Research Center for Nuclear Physics, Osaka University, Osaka 567-0047}
\affiliation{Theoretical Research Division, Nishina Center, RIKEN, Saitama 351-0198}
\affiliation{RIKEN BNL Research Center, Upton, New York 11973}
\affiliation{Saga University, Saga 840-8502}
\affiliation{University of Science and Technology of China, Hefei 230026}
\affiliation{Seoul National University, Seoul 151-742}
\affiliation{Shinshu University, Nagano 390-8621}
\affiliation{Showa Pharmaceutical University, Tokyo 194-8543}
\affiliation{Soongsil University, Seoul 156-743}
\affiliation{University of South Carolina, Columbia, South Carolina 29208}
\affiliation{Stefan Meyer Institute for Subatomic Physics, Vienna 1090}
\affiliation{Sungkyunkwan University, Suwon 440-746}
\affiliation{School of Physics, University of Sydney, New South Wales 2006}
\affiliation{Department of Physics, Faculty of Science, University of Tabuk, Tabuk 71451}
\affiliation{Tata Institute of Fundamental Research, Mumbai 400005}
\affiliation{Excellence Cluster Universe, Technische Universit\"at M\"unchen, 85748 Garching}
\affiliation{Department of Physics, Technische Universit\"at M\"unchen, 85748 Garching}
\affiliation{Toho University, Funabashi 274-8510}
\affiliation{Tohoku Gakuin University, Tagajo 985-8537}
\affiliation{Department of Physics, Tohoku University, Sendai 980-8578}
\affiliation{Earthquake Research Institute, University of Tokyo, Tokyo 113-0032}
\affiliation{Department of Physics, University of Tokyo, Tokyo 113-0033}
\affiliation{Tokyo Institute of Technology, Tokyo 152-8550}
\affiliation{Tokyo Metropolitan University, Tokyo 192-0397}
\affiliation{Tokyo University of Agriculture and Technology, Tokyo 184-8588}
\affiliation{University of Torino, 10124 Torino}
\affiliation{Toyama National College of Maritime Technology, Toyama 933-0293}
\affiliation{Utkal University, Bhubaneswar 751004}
\affiliation{Virginia Polytechnic Institute and State University, Blacksburg, Virginia 24061}
\affiliation{Wayne State University, Detroit, Michigan 48202}
\affiliation{Yamagata University, Yamagata 990-8560}
\affiliation{Yonsei University, Seoul 120-749}
  \author{A.~Abdesselam}\affiliation{Department of Physics, Faculty of Science, University of Tabuk, Tabuk 71451} 
  \author{I.~Adachi}\affiliation{High Energy Accelerator Research Organization (KEK), Tsukuba 305-0801}\affiliation{SOKENDAI (The Graduate University for Advanced Studies), Hayama 240-0193} 
  \author{K.~Adamczyk}\affiliation{H. Niewodniczanski Institute of Nuclear Physics, Krakow 31-342} 
  \author{H.~Aihara}\affiliation{Department of Physics, University of Tokyo, Tokyo 113-0033} 
  \author{S.~Al~Said}\affiliation{Department of Physics, Faculty of Science, University of Tabuk, Tabuk 71451}\affiliation{Department of Physics, Faculty of Science, King Abdulaziz University, Jeddah 21589} 
  \author{K.~Arinstein}\affiliation{Budker Institute of Nuclear Physics SB RAS, Novosibirsk 630090}\affiliation{Novosibirsk State University, Novosibirsk 630090} 
  \author{Y.~Arita}\affiliation{Graduate School of Science, Nagoya University, Nagoya 464-8602} 
  \author{D.~M.~Asner}\affiliation{Pacific Northwest National Laboratory, Richland, Washington 99352} 
  \author{T.~Aso}\affiliation{Toyama National College of Maritime Technology, Toyama 933-0293} 
  \author{H.~Atmacan}\affiliation{Middle East Technical University, 06531 Ankara} 
% ...
\collaboration{The Belle Collaboration}

\noaffiliation
\begin{abstract}

We report the first observation of the polarization of $\Lambda/\bar{\Lambda}$ hyperons transverse to its production plane in $e^+e^-$ annihilation. We observe a significant polarization that rises with the fractional energy carried by the hyperon as well as its transverse momentum. To define the production plane, we use the direction of the hyperon momentum together with either the thrust axis in the event or the momentum vector of a hadron in the opposite hemisphere. Furthermore, we investigate the contributions to the hyperon polarization from the feed-down from $\Sigma^0/\bar{\Sigma}^0$ and $\Lambda_c^{\pm}$ decays.
This measurement uses a dataset of 800.4~fb$^{-1}$ collected by the Belle experiment at or near a center-of-mass energy of 10.58 GeV.

\end{abstract}


\maketitle

\tighten

{\renewcommand{\thefootnote}{\fnsymbol{footnote}}}
\setcounter{footnote}{0}
Due to its self-analyzing weak decay, the $\Lambda/\bar{\Lambda}$ hyperon (in the following denoted by $\Lambda$) has played a key role in studying the transverse spin structure of hadrons. Historically, the observation of large transverse $\Lambda$ polarization in unpolarized $pp$ collisions has been one of the key measurements that motivated a successful program to investigate transverse spin effects in nuclear physics~(see Refs.~\cite{Panagiotou:1989sv,Metz:2016swz} and references therein). However, at present time, the precise mechanism behind the $\Lambda$ polarization remains unknown. 
It is assumed that processes related to the polarizing fragmentation function $D_{1T}^{\perp \Lambda/q}(z, p^2_{\perp})$ play a major part in these effects.
Here, $z$ denotes the fractional energy of the fragmenting quark carried by the observed hadron and $p_{\perp}$ denotes its transverse momentum relative to the fragmenting quark. In addition to shedding light on the observed large polarization in hadroproduction of $\Lambda$'s, $D_{1T}^{\perp \Lambda/q}$ is of particular interest since, at leading twist and for non-perturbative intrinsic transverse momenta, it represents the fragmentation analog to the Sivers parton distribution function $f_{1T}^{\perp}(x,k_{t})$~\cite{Pitonyak:2013dsu}.
While $f_{1T}^{\perp}(x,k_{t})$ describes the transverse momentum ($k_{t}$) dependence of partons (carrying a fraction $x$ of the nucleon momentum) inside a polarized nucleon, $D_{1T}^{\perp \Lambda/q}$ describes the transverse momentum dependence of a fragmenting polarized hyperon. The Sivers function has been the subject of intense theoretical and experimental interest due to its connection to quark orbital angular momentum, a missing piece of the spin puzzle of the nucleon, and fundamental predictions rooted in the gauge invariance of QCD pertaining to a sign change~\cite{Brodsky:2002cx,Collins:2002kn,Ji:2002aa,Belitsky:2002sm} of $f_{1T}^{\perp}$ in hadronic collisions compared to the one observed in semi-inclusive deep inelastic scattering~\cite{Airapetian:2009ae}. Similarly, a non-vanishing $D_{1T}^{\perp \Lambda/q}$ can help to shed light on the spin structure of the $\Lambda$ and has been proposed as a fundamental test of universality between different processes of similar importance as the Sivers sign change~\cite{Boer:2010ya}.
To our knowledge, a nonzero effect of $D_{1T}^{\perp \Lambda/q}$ has not been observed in $e^+e^-$ annihilation. The OPAL experiment at LEP looked for transverse $\Lambda$ polarization~\cite{Ackerstaff:1997nh}, but no significant signal was observed. It is thought that the absence of a signal may be attributed to the high center-of-mass energy $\sqrt{s}$, which increases the phase space for gluon radiation that, in turn, can dilute the effect, and low statistics compared to the $B$-factories.

Because parton distribution functions do not enter in the cross-section, $e^+e^-$ annihilation is uniquely suited to access fragmentation functions~\cite{Metz:2016swz}. We present here the first observation of the transverse polarization of $\Lambda$ produced in $e^+e^-$ annihilation, from which $D_{1T}^{\perp \Lambda/q}$ can be extracted.
We use a dataset of 800.4~fb$^{-1}$ collected by the Belle experiment at or near $\sqrt{s}= 10.58$ GeV.
The Belle instrumentation~\cite{BelleDetector} at the KEKB $e^+e^-$ collider~\cite{KEKB} used in this analysis includes the central drift chamber (CDC) and the silicon vertex detector, which provides precision tracking for charged particles, 
and the electromagnetic calorimeters (ECL). 
Particle identification (PID) is performed using information on $dE/dx$ in the CDC, a time-of-flight system in the barrel, aerogel Cherenkov counters in the barrel and the forward endcap, as well as a muon and $K_L$ identification system outside the superconducting solenoid, which provides a 1.5~T magnetic field. 
To correct the data for detector effects and for systematic studies, Monte Carlo (MC) simulated events are generated by
Pythia6.2~\cite{Sjostrand:2000wi}
and processed with a GEANT3~\cite{Brun:1987ma}-based full simulation of the Belle detector. 

This measurement considers the inclusive processes $e^+e^-\rightarrow \Lambda + X$ as well as $e^+e^-\rightarrow \Lambda + h^{\pm} + X$, where $h$ denotes a light hadron detected in the hemisphere opposite the $\Lambda$; the hemispheres are determined using the thrust axis in the event. The primary reason to detect another hadron in the opposite hemisphere is to provide additional information on the flavor of the parent $q\bar{q}$ pair.
For instance, in the process $e^+e^-\rightarrow \Lambda + \pi^{-} + X$, one can assume the leading contribution comes from $u$ quark fragmenting into $\Lambda$ and $\bar{u}$ fragmenting into $\pi^{-}$~\cite{Boer:2010ya}. 
In this analysis, we consider charged pions and kaons $(\pi^{\pm}, K^{\pm})$ in the opposite hemisphere.

Using the event-shape-sensitive thrust $T$, a sample of light and charm quark ($u,d,s,c$) fragmentation is selected~\cite{Seidl:2008xc, Vossen:2011fk}. 
The thrust is defined in the center-of-mass of $e^+e^-$ system as 
\begin{equation}
T= \max_{|\vec{T}|=1} \frac{\sum_i |\vec{T}\cdot \vec{p}_i|}{\sum_i |\vec{p}_i|}.
\end{equation}
Here, the $\vec{p}_i$ are the momenta of all particles in the event and $\vec{T}$ is the normalized thrust direction. We require of $T>0.8$, which reduces the contribution of $\Upsilon$ decays to less than 1\%. 

\begin{figure}
\begin{center}
\includegraphics[width=0.5\textwidth]{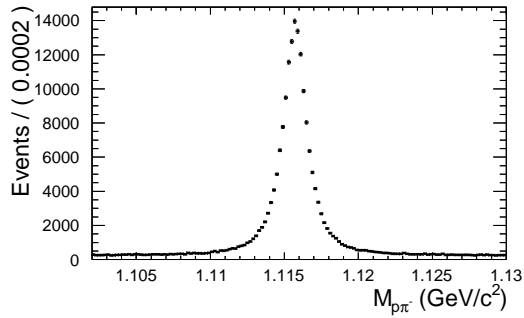}
\caption{The distribution of invariant mass $M_{p\pi^-}$. 
The background fraction is pretty small.
We use the sidebands to correct for the background under the signal peak, as explained in the text. 
\label{fig:Lmass}}
\end{center}
\end{figure}

In each event, we reconstruct $\Lambda$ ($\bar{\Lambda}$) candidates in the channel $\Lambda \rightarrow p + \pi^{-}$ ($\bar{\Lambda} \rightarrow \bar{p} + \pi^{+}$). 
The $\Lambda$ candidate is required to have a displaced vertex, consistent with a long-lived particle originating
from the IP. The daughter proton and pion are constrained to a decay vertex and the four-momenta are refitted with the vertex constraint. To further suppress backgrounds, we require the likelihood ($\mathcal{L}$) calculated using the Belle PID detectors for one of the daughter particle to be a proton ($p$) to be $\mathcal{L}(p)>$0.6, where the efficiency of this selection is about 90\%. No such requirement is imposed on the pion candidate. The distribution of invariant mass $M_{p\pi^-}$ is shown in Fig.~\ref{fig:Lmass}, where the signal peak is quite clear and the sidebands are used to correct for the remaining background under the signal peak.
All other charged tracks in the event, with the exception of the $\Lambda$ daughter particles, are required to have an impact parameter with respect to the interaction point of less than 2.0\,cm in the $r-\phi$ plane and 4.0\,cm along the $z$ axis. A likelihood ratio $\mathcal{L}(K)$/$\mathcal{L}(\pi)$ is required to select the light hadrons in the opposite hemisphere. It must be larger than 0.6 to identify $K^\pm$, 
where the efficiency of kaon exceeds 80\% and pion fake rate is below 10\%. 
The likelihood ratio must be less than 0.4 to identify $\pi^\pm$; the pion efficiency is about 90\% and the kaon fake rate is below 10\%.

To determine the transverse momentum of the hyperon and the event plane with respect to which the transverse momentum of the hadron is measured, one more direction is needed. In the elementary scattering, the second axis is given by the parent $q\bar{q}$ axis, which is not observed. Here, we use two alternative frames. One uses the thrust axis as a proxy for the $q\bar{q}$ axis and the other uses the axis of the detected hadron in the opposite hemisphere. We will refer to these as the ``thrust frame" and the ``hadron frame", respectively. In the thrust frame alone, an inclusive $\Lambda$ measurement is possible; in the hadron frame, all combinations of a $\Lambda$($\bar{\Lambda}$) and a light hadron are considered. 
In the thrust frame, the thrust axis direction is chosen so that it points into the same hemisphere as the $\Lambda$ and; in the hadron frame, the hadron's momentum is reversed to also point into this hemisphere. 
We call the defining axis ($\vec{T}$ or -$\hat{p}_h$) as $\vec{m}$, then define the direction $\hat{n} \propto \vec{m} \times \hat{p}_{\Lambda}$, the transverse polarization of $\Lambda$ is investigated along $\hat{n}$. Given a polarization $P$ of the $\Lambda$, the decay distribution is
\begin{equation}
\label{eq:main}
\frac{1}{N} \frac{dN}{d \rm{cos}\theta} = 1 + \alpha P {\rm cos}\theta,
\end{equation}
where $N$ is the signal yield, $\theta$ is the angle between $\hat{n}$ and the proton momentum in the $\Lambda$ rest frame, and $\alpha$ is the decay parameter of the decay process $\Lambda \rightarrow p + \pi^{-}$ ($\bar{\Lambda} \rightarrow \bar{p} + \pi^{+}$). The world average of $\alpha$ is 0.642 $\pm$ 0.013 for $\Lambda$ and $-$0.71 $\pm$ 0.08 for $\bar{\Lambda}$~\cite{PDG}. 

\begin{figure}
\begin{center}
\includegraphics[width=0.8\textwidth]{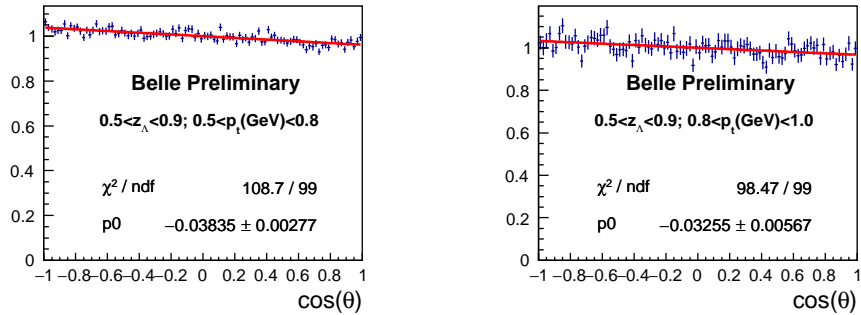}
\caption{The fits to the efficiency-corrected $\cost$ distributions in two bins: $z_{\Lambda}$=[0.5, 0.9]; $p_{\rm t}$=[0.5, 0.8]\,GeV (left) and $z_{\Lambda}$=[0.5, 0.9]; $p_{\rm t}$=[0.8, 1.0]\,GeV (right). 
\label{fig:costheta_fit}}
\end{center}
\end{figure}

The signal region for the reconstructed $\Lambda$ mass is chosen as [1.11, 1.12]\,GeV/$c^2$ and the lower and upper sidebands as [1.103, 1.108]\,GeV/$c^2$ and [1.123, 1.128]\,GeV/$c^2$, respectively. 
The $\cos\theta$ distribution in the sideband region is subtracted from that in the signal region.
Since the background shape is basically flat and we choose a common width for signal and sideband regions, the sideband scale factor in this subtraction is 1. 
The subtracted distribution is self-normalized: $R(\theta) = N(\theta)/{\langle N \rangle}$,
where $\langle N \rangle$ denotes the averaged number of events in each bin. The resulting yield is corrected with the efficiencies obtained from the simulation. 
To avoid the efficiency corrections, we also construct the ratio between the $\cost$ distributions of $\Lambda$ and $\bar{\Lambda}$. In this ratio, the efficiencies cancel if one assumes that they do not depend on the charge of the detected particle.
The sideband-subtracted and efficiency-corrected $\cost$ distribution is fitted using the function $1 + p_{0} {\rm cos}\theta$,
where $p_{0}$ is a free parameter. The goodness of fit indicates that the data is described well by this functional form, as shown in Fig.~\ref{fig:costheta_fit}.
The magnitude of the polarization is then $P= p_{0}/\alpha$. 

\begin{figure}
\begin{center}
\includegraphics[width=0.65\textwidth]{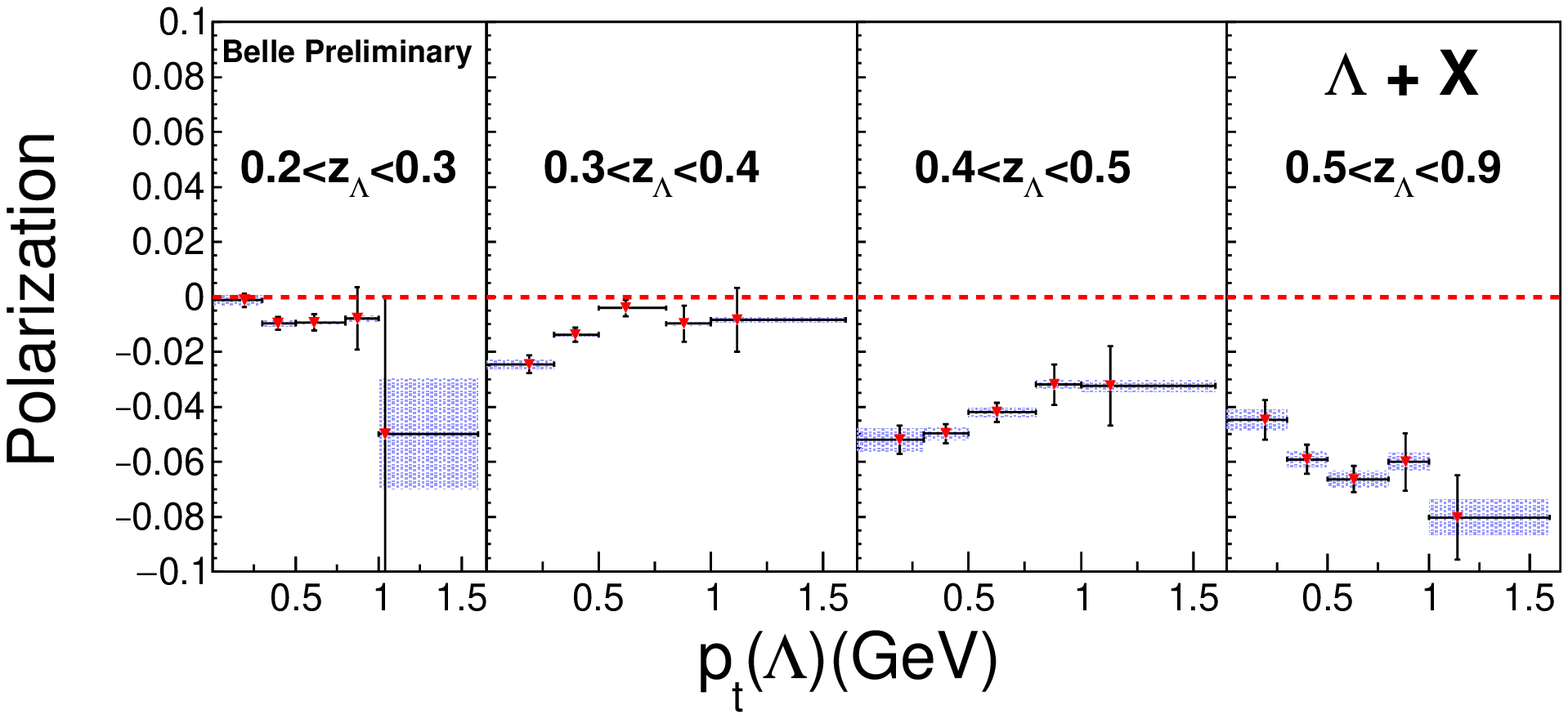}
\includegraphics[width=0.65\textwidth]{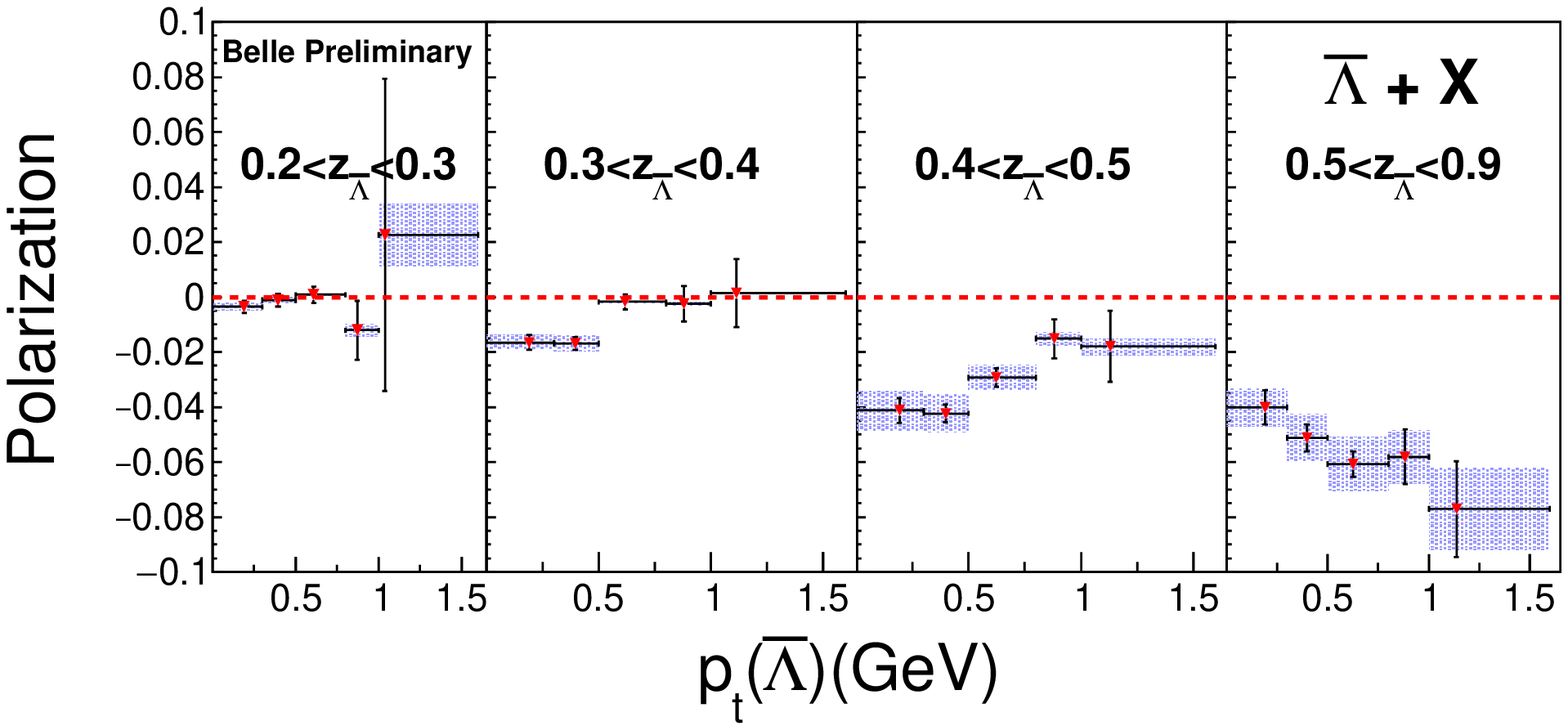}
\includegraphics[width=0.65\textwidth]{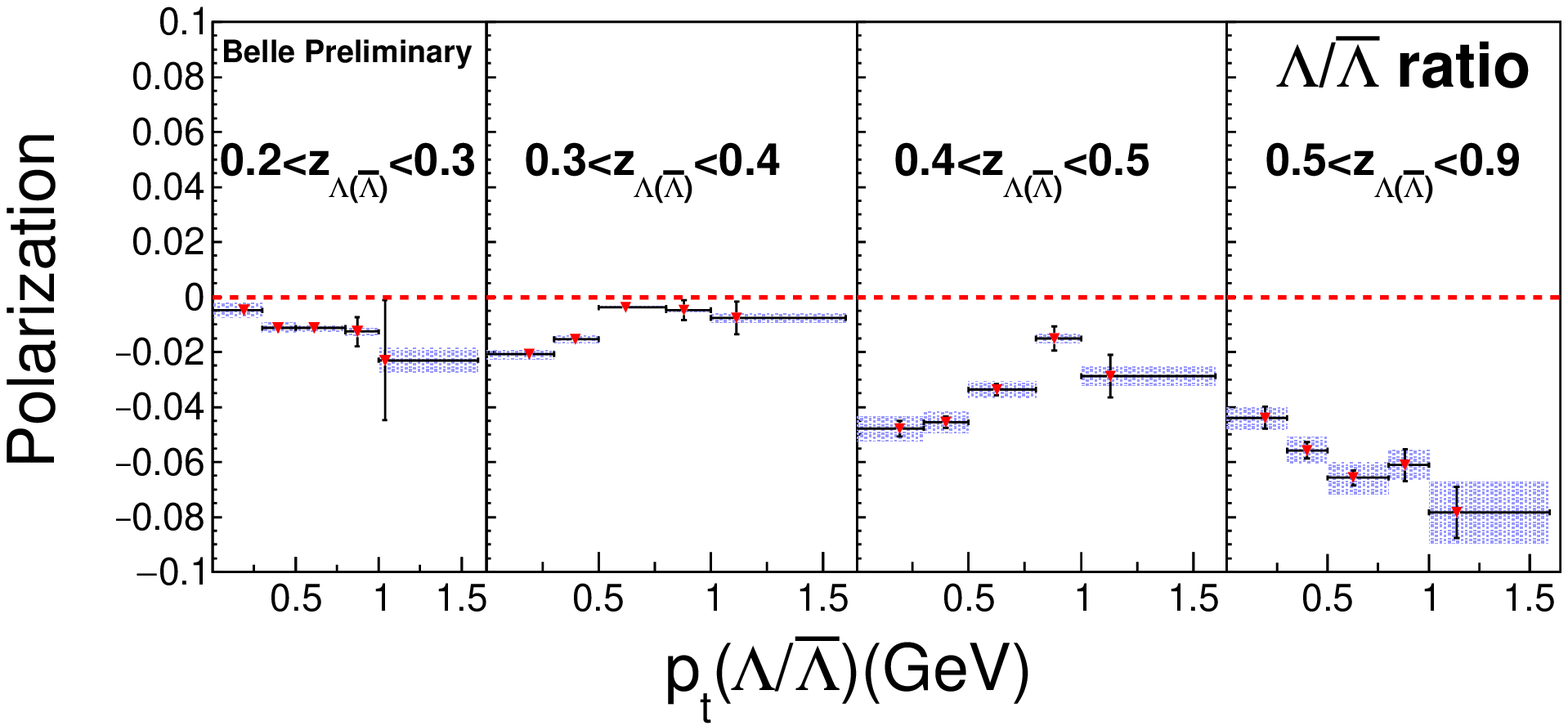}
\caption{Transverse polarization amplitudes of inclusive $\Lambda$ as a function of $z_\Lambda$ and $p_{\rm t}$ in the thrust frame. The error bars show statistical uncertainties and the shaded areas show the systematic uncertainties. The first and second panels show results for $\Lambda$ and $\bar{\Lambda}$, respectively, using efficiency from MC as efficiency shape. The bottom panel shows the results from the $\Lambda$-$\bar{\Lambda}$ ratios, where the efficiency uncertainties cancel.
\label{fig:pol_vszpt_sys}}
\end{center}
\end{figure}

The transverse polarization of the $\Lambda$ is investigated as a function of $z_{\Lambda}$ and $p_{\rm t}$, where $z_{\Lambda} = 2E_{\Lambda}/\sqrt{s}$ and $p_{\rm t}$ is the transverse momentum of the $\Lambda$ with respect to the thrust axis (hadron momentum) in the thrust frame (hadron frame). Four $z_\Lambda$ bins with boundaries at $z_{\Lambda} = [0.2, 0.3, 0.4, 0.5, 0.9]$ and five $p_{\rm t}$ bins with boundaries at $p_{\rm t} = [0.0, 0.3, 0.5, 0.8, 1.0, 1.6]\,\rm GeV$ are adopted in the thrust frame. 
Results are summarized in Fig.~\ref{fig:pol_vszpt_sys} with error bars showing the statistical uncertainties;  the shaded areas showing the systematic uncertainties are discussed later.
A clear nonzero transverse polarization is observed. In general, this polarization rises with $z_\Lambda$. The $p_{\rm t}$ behavior is more complex and depends on the $z_\Lambda$ range. For $z_\Lambda>$ 0.5 where the $\Lambda$ is the leading particle and for $z_\Lambda <$0.3, we observe rising asymmetries with $p_{\rm t}$. In contrast, for intermediate $z_\Lambda$, the dependence reverses with a possible minimum around 1~GeV. The observed amplitudes are consistent between $\Lambda$ and $\bar{\Lambda}$.

\begin{figure}
\begin{center}
\includegraphics[width=0.48\textwidth]{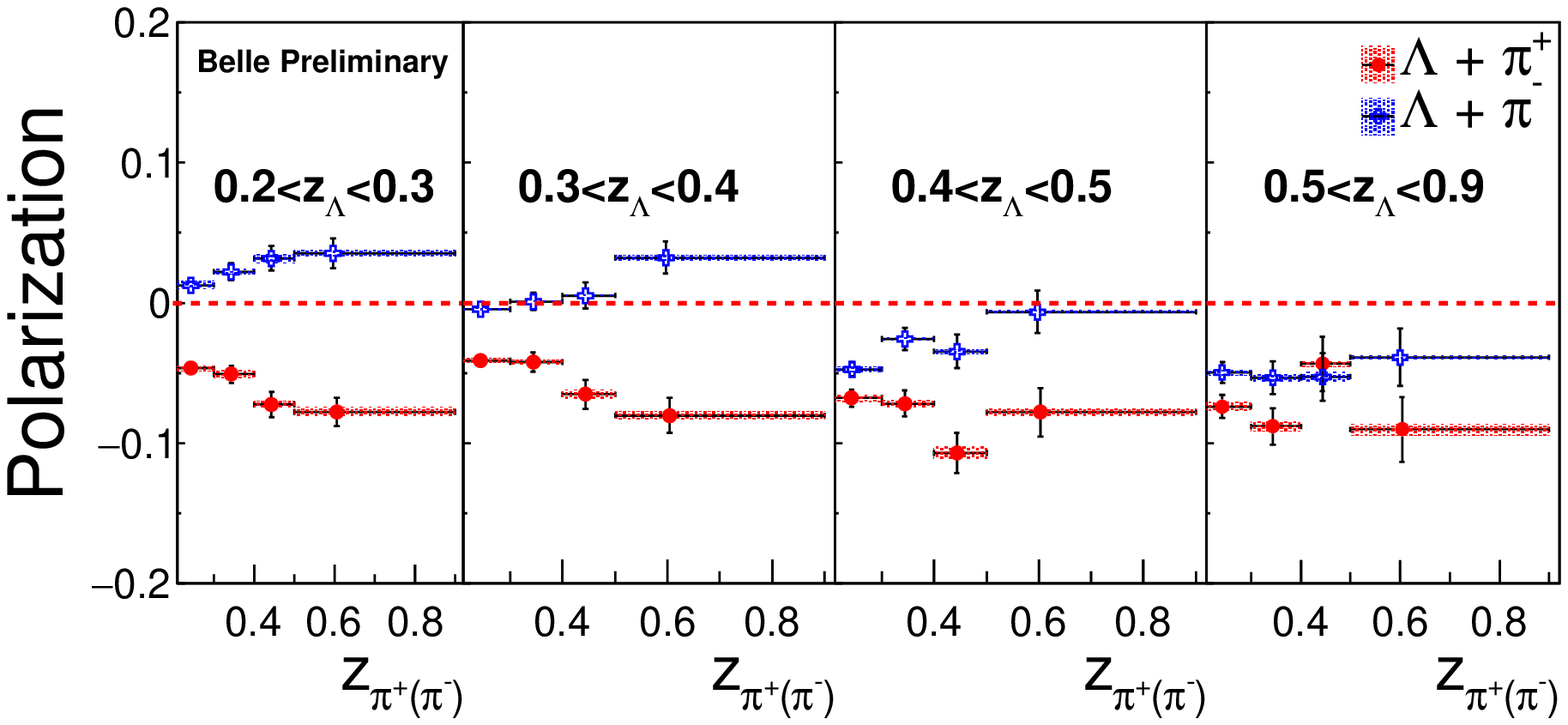}
\includegraphics[width=0.48\textwidth]{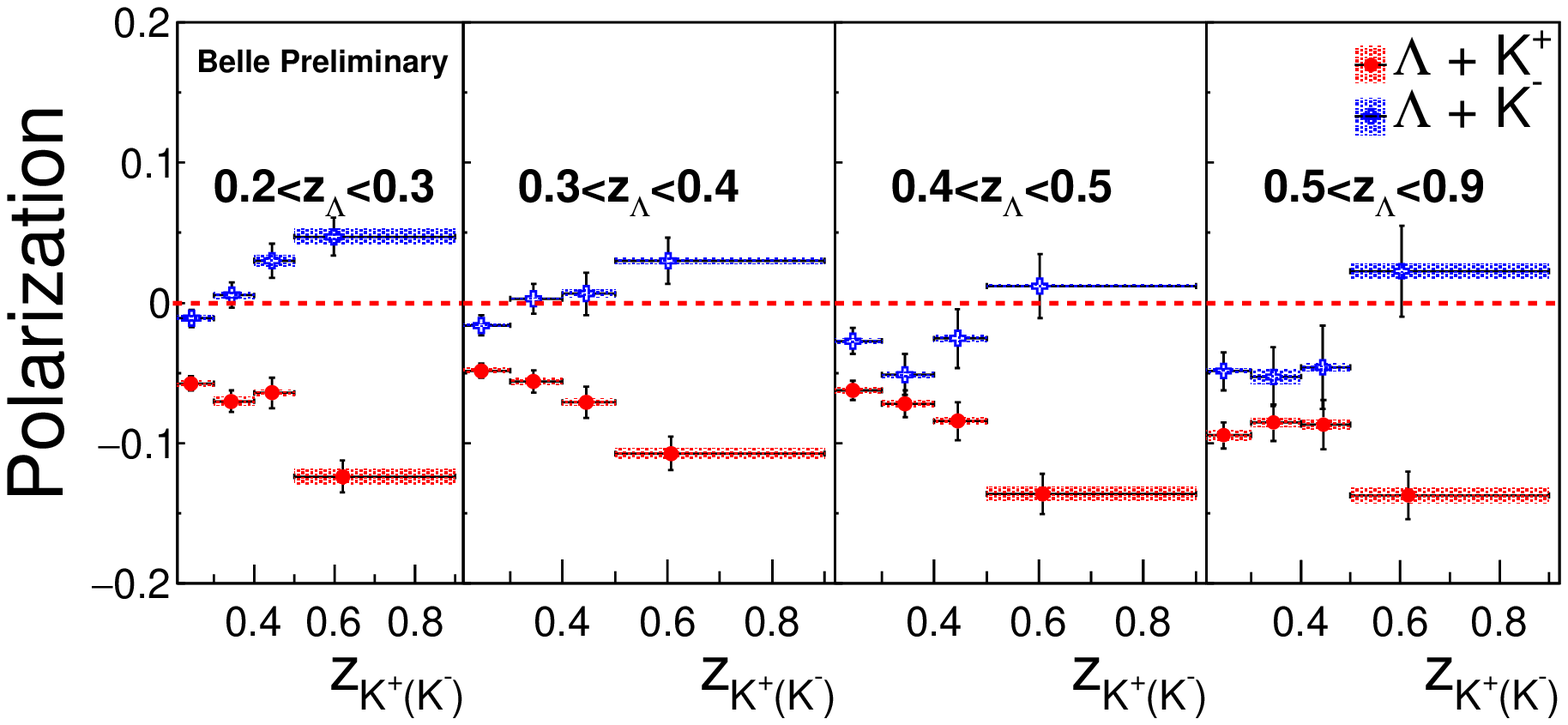}
\includegraphics[width=0.48\textwidth]{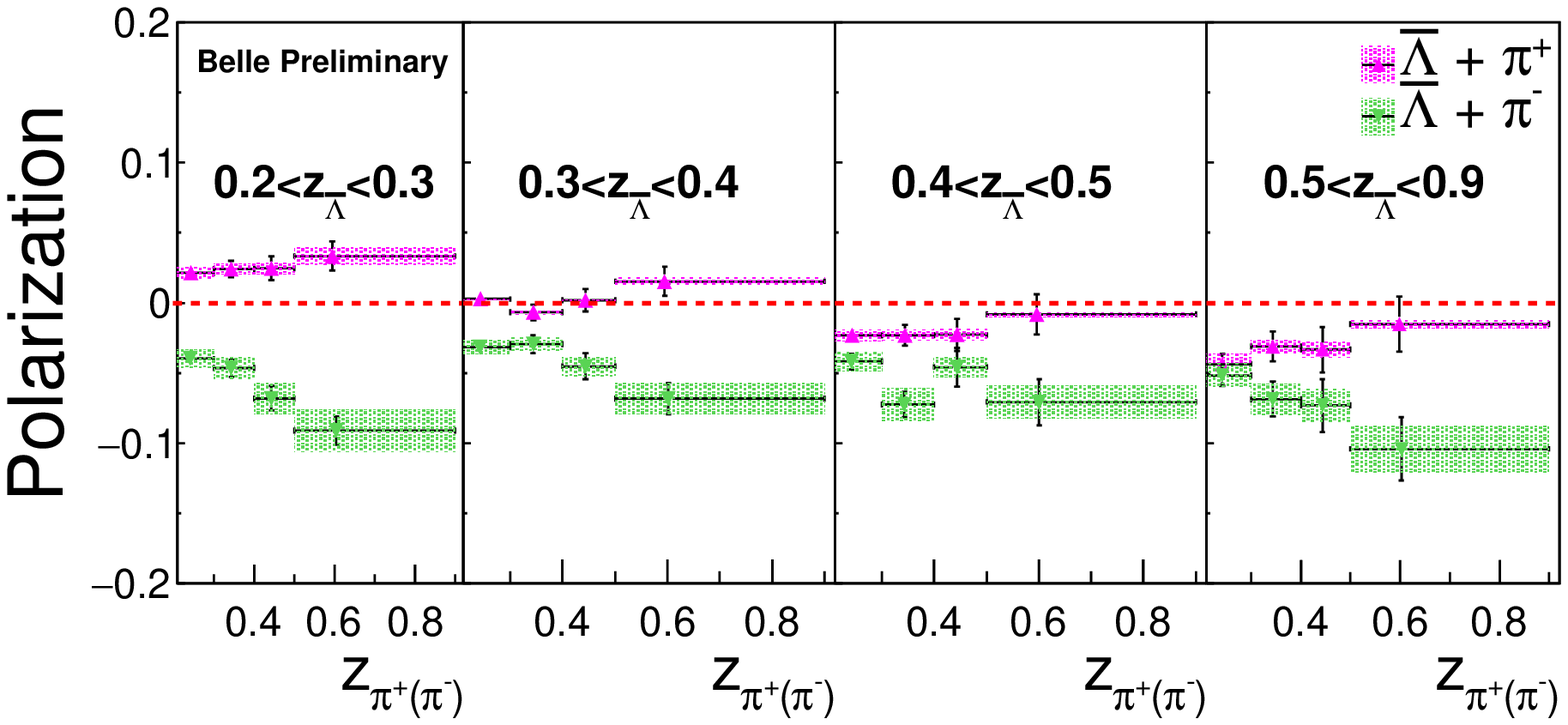}
\includegraphics[width=0.48\textwidth]{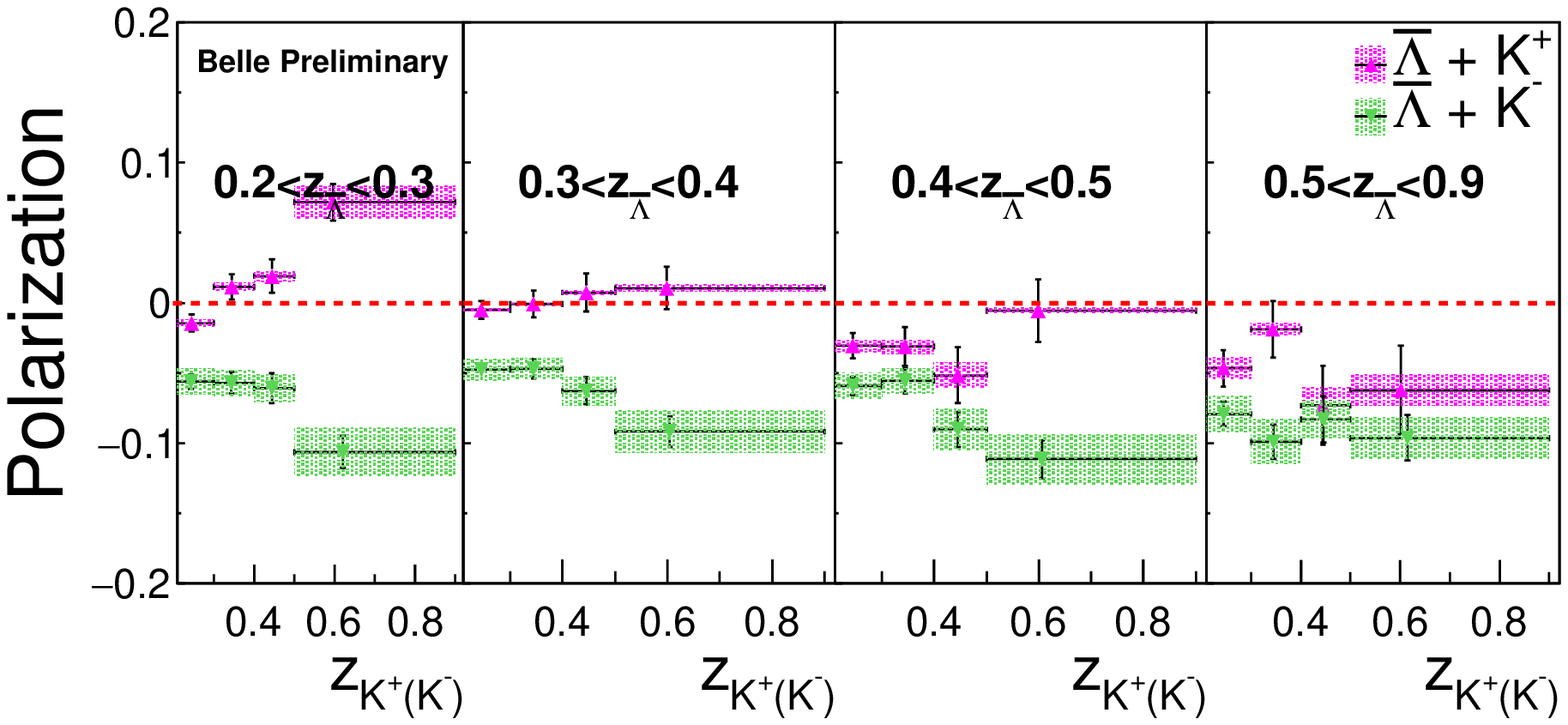}
\caption{Transverse polarizations of $\Lambda$ observed in $\Lambda + h^{\pm} + X $ as a function of $z_\Lambda$ and $z_{h}$ in the thrust frame. The error bars show statistical uncertainties and the shaded areas show the systematic uncertainties.}
\label{fig:pol_zlzh_sys}
\end{center}
\end{figure}

\begin{figure}
\begin{center}
\includegraphics[width=0.48\textwidth]{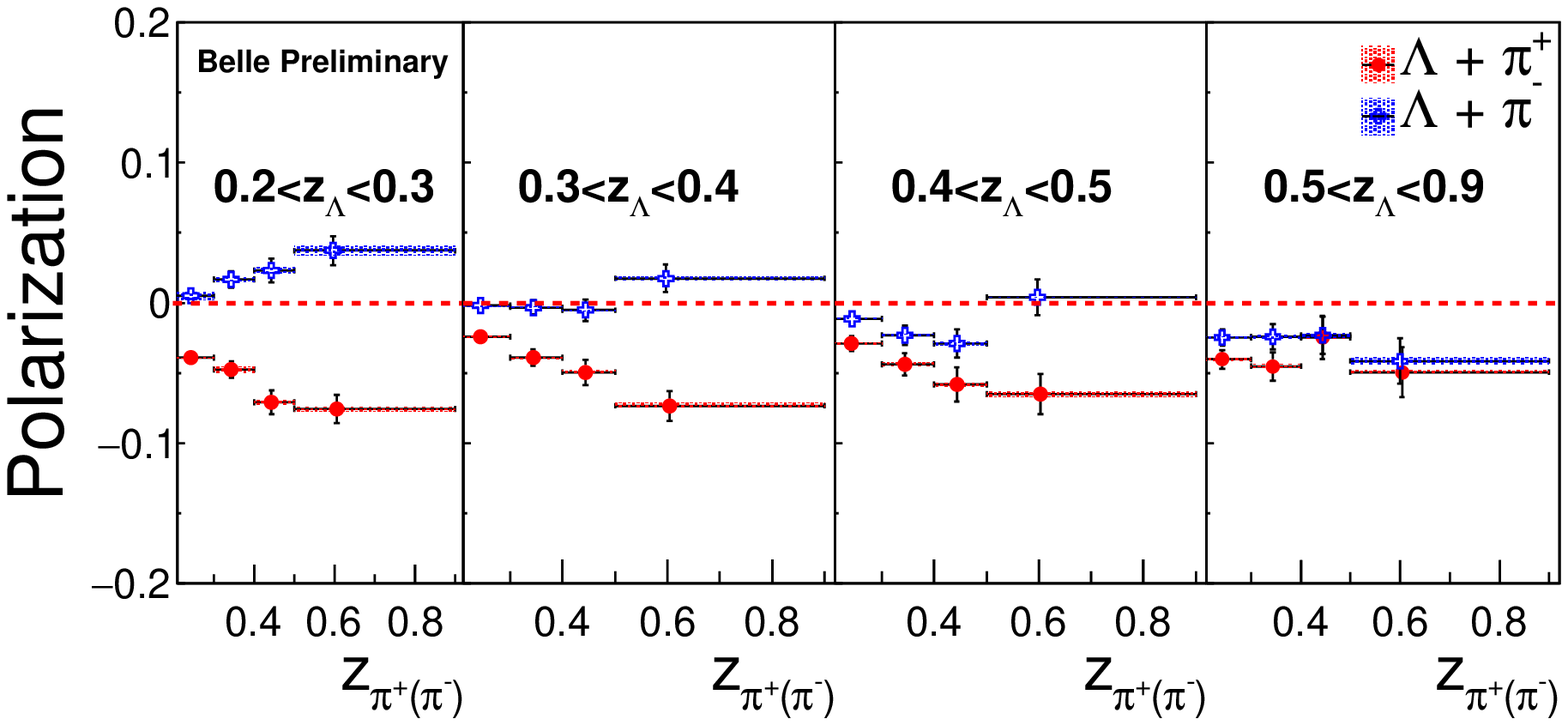}
\includegraphics[width=0.48\textwidth]{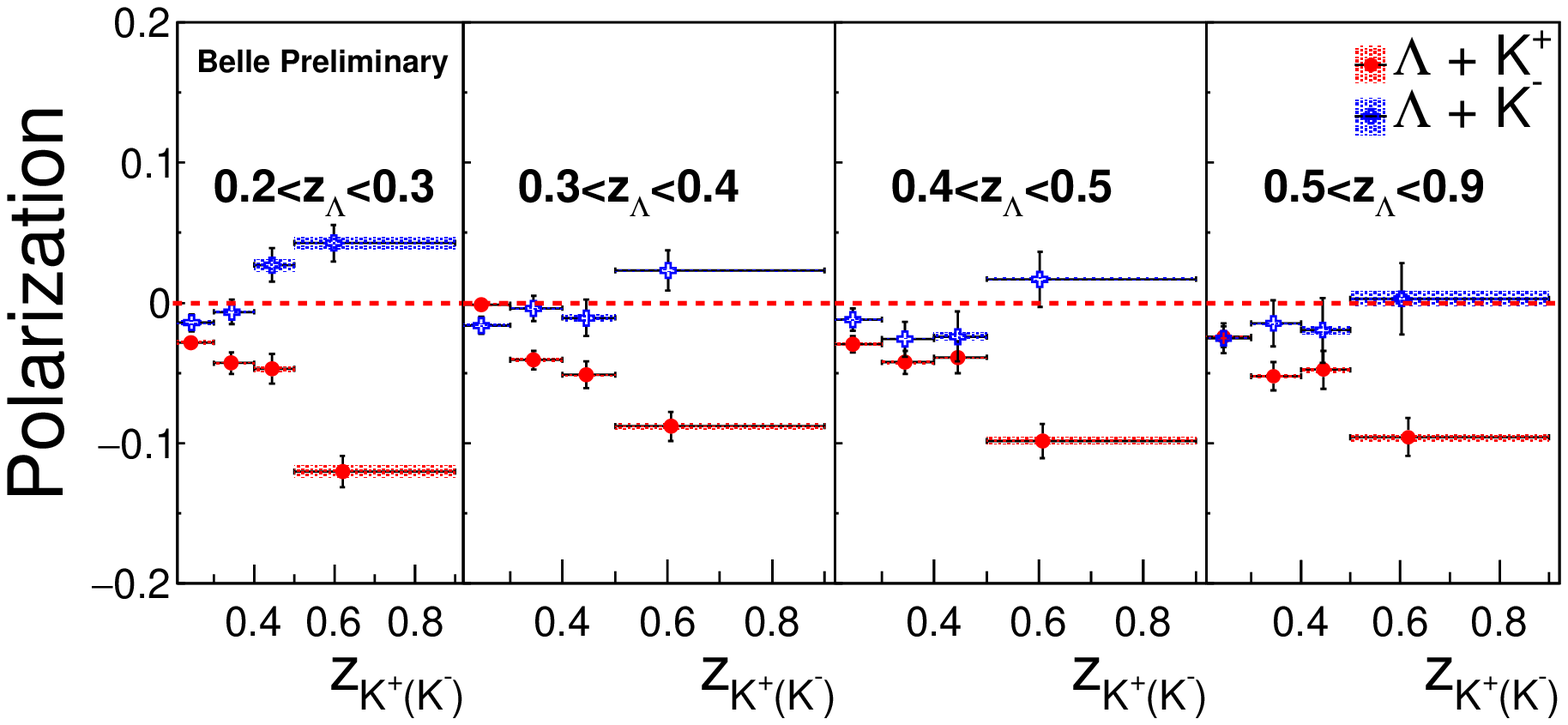}
\includegraphics[width=0.48\textwidth]{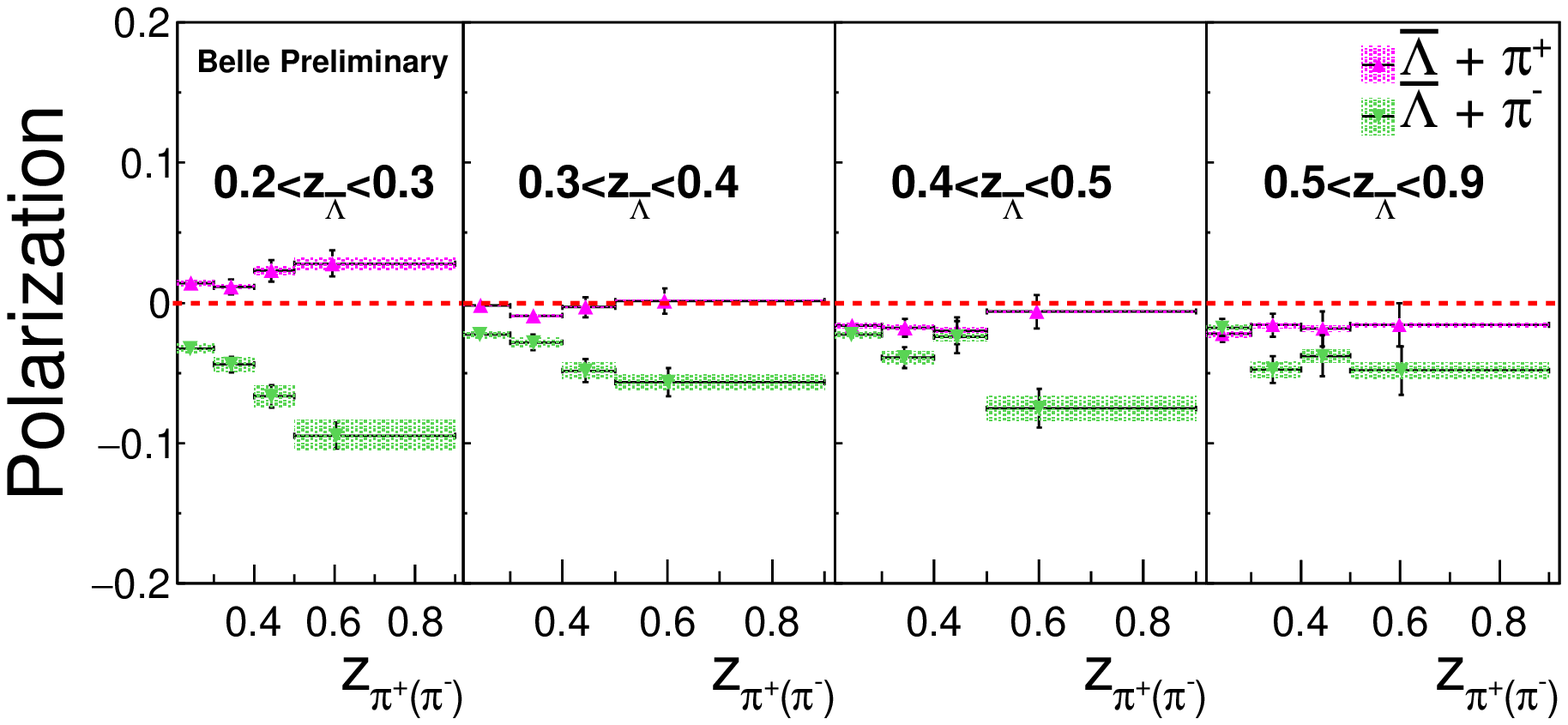}
\includegraphics[width=0.48\textwidth]{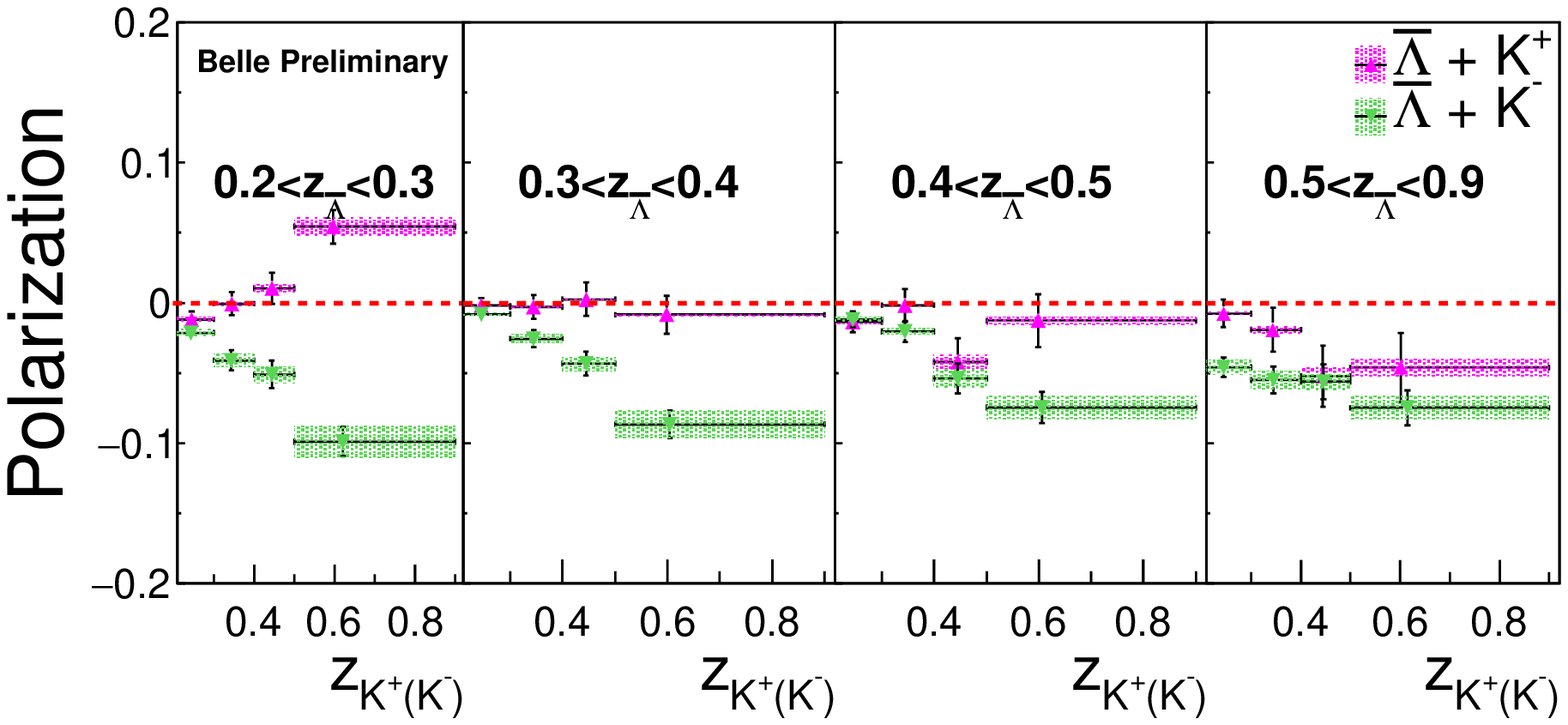}
\caption{Transverse polarizations of $\Lambda$ observed in $\Lambda + h^{\pm} + X $ as a function of $z_\Lambda$ and  $z_{h}$ in the hadron frame. The error bars show statistical uncertainties and the shaded areas show the systematic uncertainties.}
\label{fig:pol_zlzh_sys_href}
\end{center}
\end{figure}

When considering a light hadron in the opposite side, four $z_{h}$ bins with boundaries at $z_{h} = [0.2, 0.3, 0.4, 0.5, 0.9]$ are adopted; here, the $z_{h}$ is the fractional energy carried by the light hadron. Transverse polarization amplitudes of $\Lambda$ as a function of $z_{\Lambda}$ and $z_{h}$ are shown in Fig.~\ref{fig:pol_zlzh_sys} for the thrust frame and in Fig.~\ref{fig:pol_zlzh_sys_href} for the hadron frame. 
We observe similar polarizations in frames.
The flavor contribution can be revealed through different combinations of $\Lambda$ and light hadron types and charges. In particular, in the low $z_\Lambda$ region, the polarization in $\Lambda + h^+ $ and $\Lambda + h^- $ processes is significantly different, even showing opposite sign and a magnitude that increases with higher $z_h$. In contrast, in the high $z_\Lambda$ region, 
the differences between $\Lambda + h^+$ and $\Lambda + h^-$ are modest, although deviations can still be seen.

\begin{figure}
\begin{center}
\includegraphics[width=0.4\textwidth]{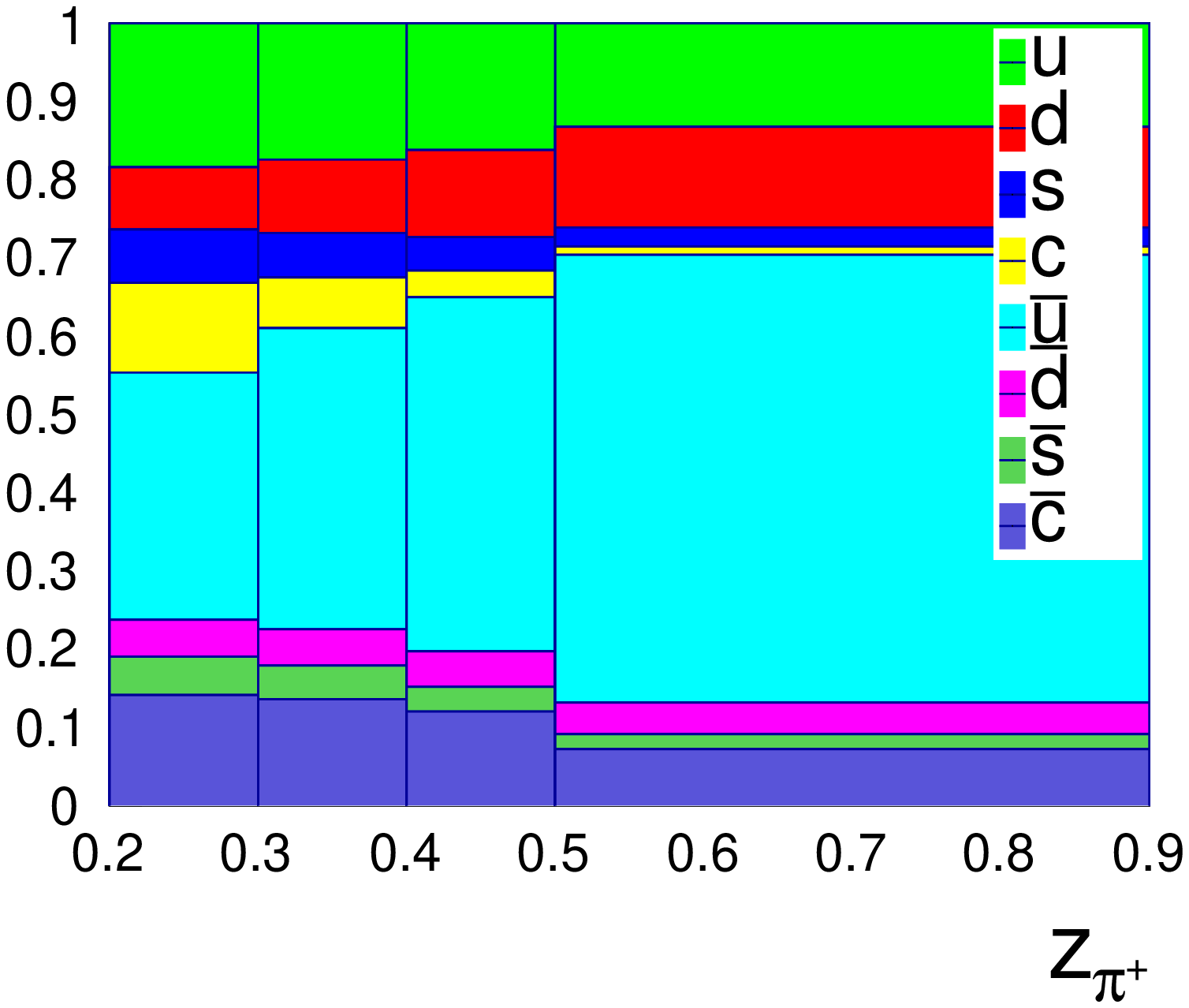}
\includegraphics[width=0.4\textwidth]{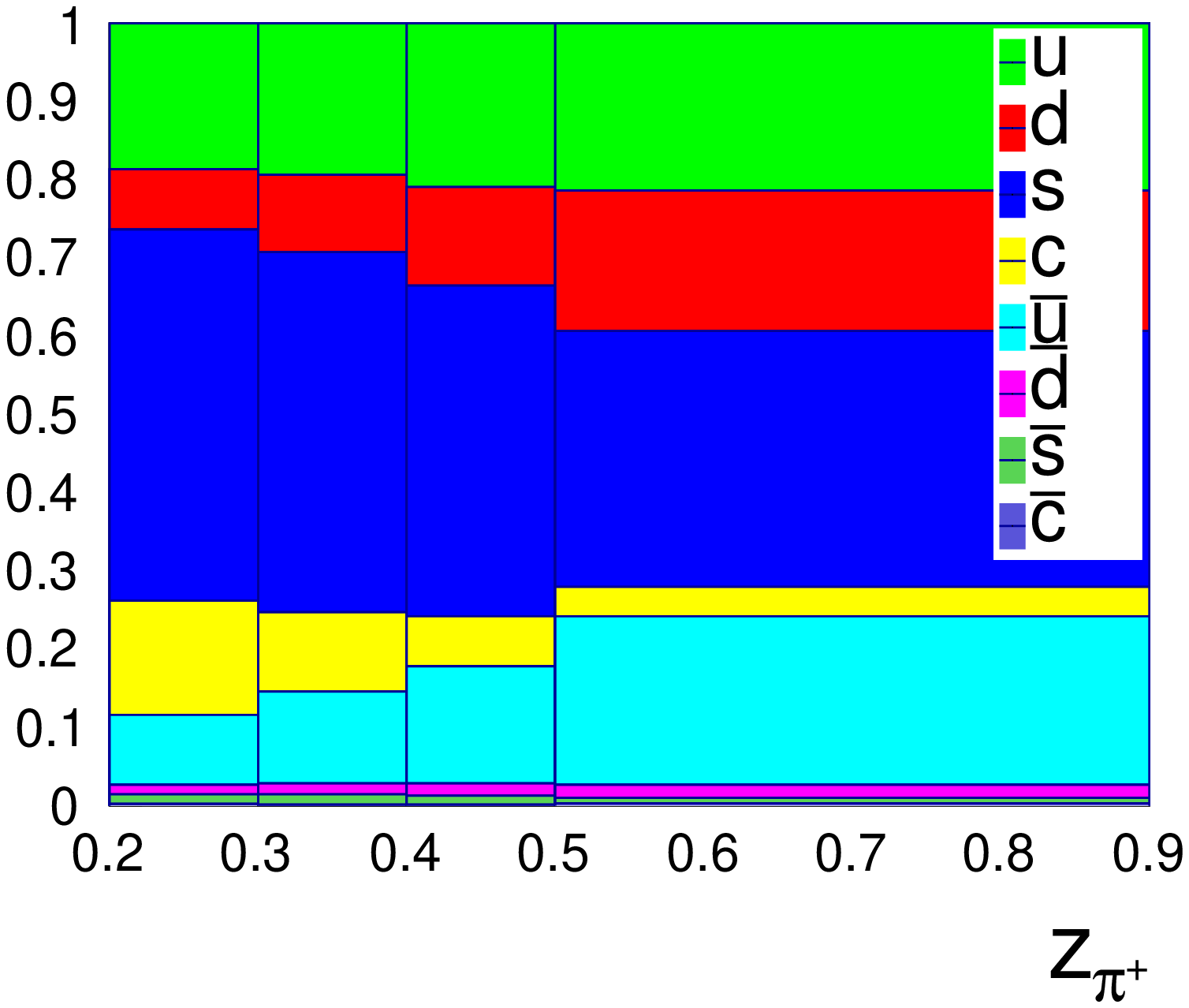}
\includegraphics[width=0.4\textwidth]{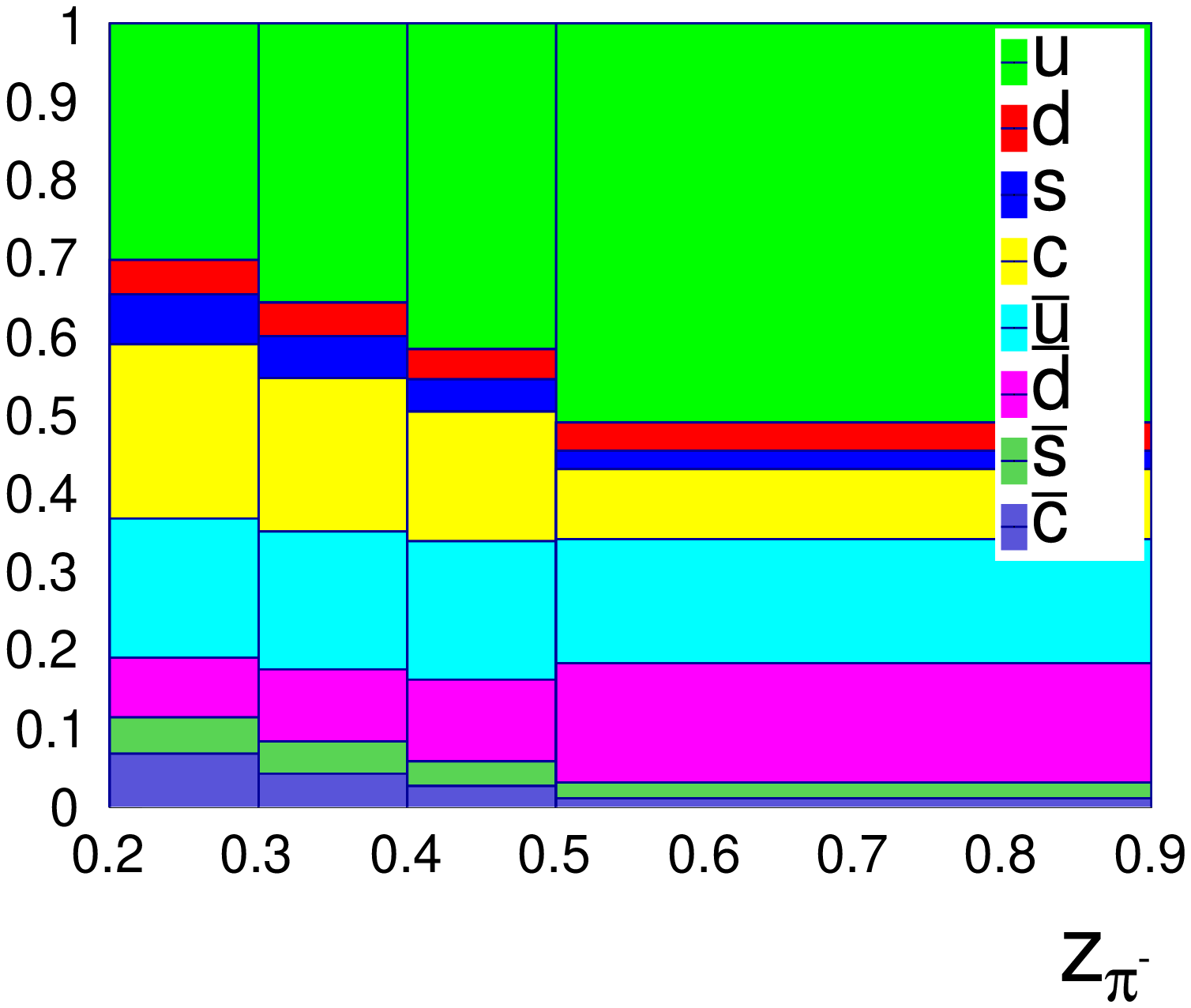}
\includegraphics[width=0.4\textwidth]{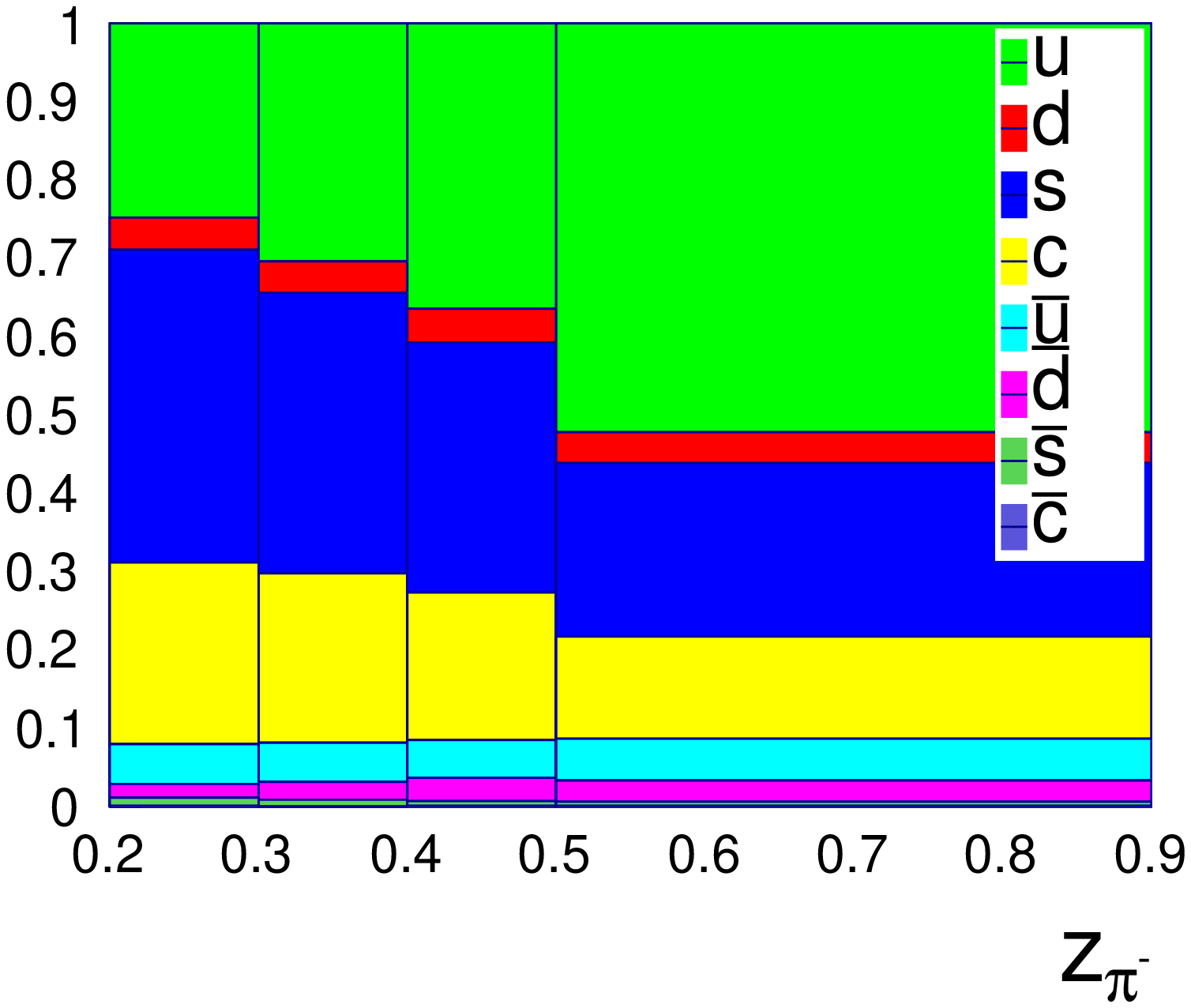}
\caption{The fractions of various quarks which go to the $\Lambda$'s hemisphere in the inclusive process $\Lambda + \pi^{+} + X $ (top) and $\Lambda + \pi^{-} + X $ (bottom) in different $z_{h}$ region at $z_\Lambda$=[0.2, 0.3] (left) and $z_\Lambda$=[0.5, 0.9] (right).}
\label{fig:flavor_tag}
\end{center}
\end{figure}

We make a simple attempt to explain the data results together with knowledge from MC. 
In MC, we search for the (anti-)quark going into the same hemisphere as the $\Lambda$.
The fractions of various quark flavors in the low-$z_\Lambda$ ([0.2, 0.3]) and high-$z_\Lambda$ ([0.5,0.9]) region for the $\Lambda + \pi^{\pm} + X$ processes are shown in Fig.~\ref{fig:flavor_tag}. 
We see that the flavor tag of the light hadron is more effective at low $z_\Lambda$ and high $z_h$. 
In this region, for $\Lambda + \pi^+ + X$, the $\bar{u} \rightarrow \Lambda(uds)$ process dominates because the $u$ is more likely to be carried away by $\pi^+(u\bar{d})$.
For $\Lambda + \pi^- + X$, $u \rightarrow \Lambda(uds); \bar{u}\rightarrow \pi^-(\bar{u}d)$ dominates. 
This might explain why the polarizations of $\Lambda + h^+ + X $ and $\Lambda + h^- + X $ have opposite sign at low $z_\Lambda$ and high $z_h$. It should be noted that (anti-)charm fragmentation contributes also, particularly at low $z_\Lambda$ in the $\Lambda + \pi^{\pm} +X$ sample.
When $z_\Lambda$ is sufficiently high, especially for low $z_h$, the flavor-tag effect is reduced. In $\Lambda + \pi^+ + X$, the $\bar{u} \rightarrow \Lambda$ contribution decreases significantly and the contribution from $s$ quarks increases in both $\pi^+$ and $\pi^-$ cases when compared to the low-$z_\Lambda$ region. Therefore, the differences in polarization between the $\Lambda + h^+ + X $ and $\Lambda + h^- + X$ samples are smaller. 

\begin{figure}
\begin{center}
\includegraphics[width=0.48\textwidth]{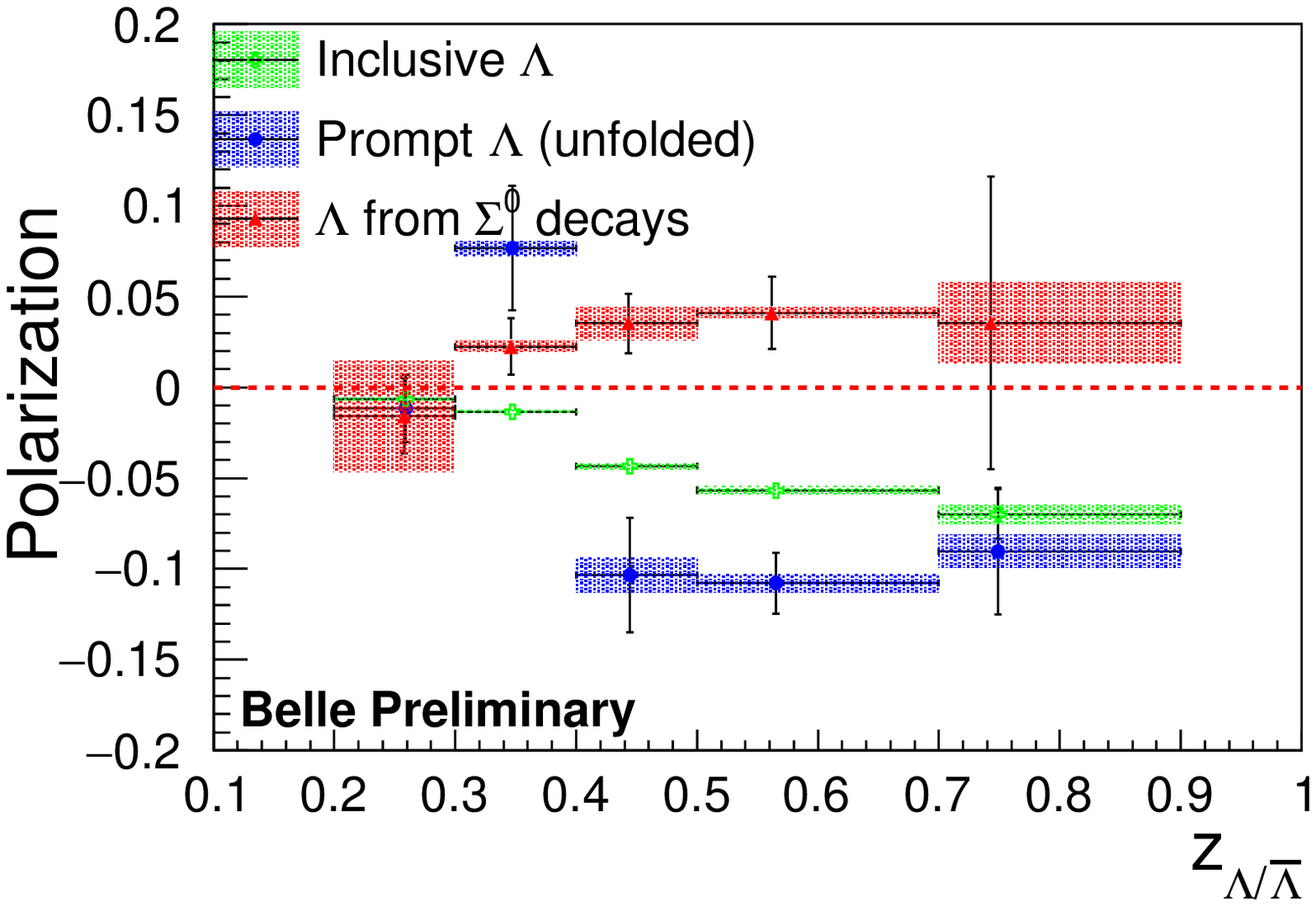}
\includegraphics[width=0.48\textwidth]{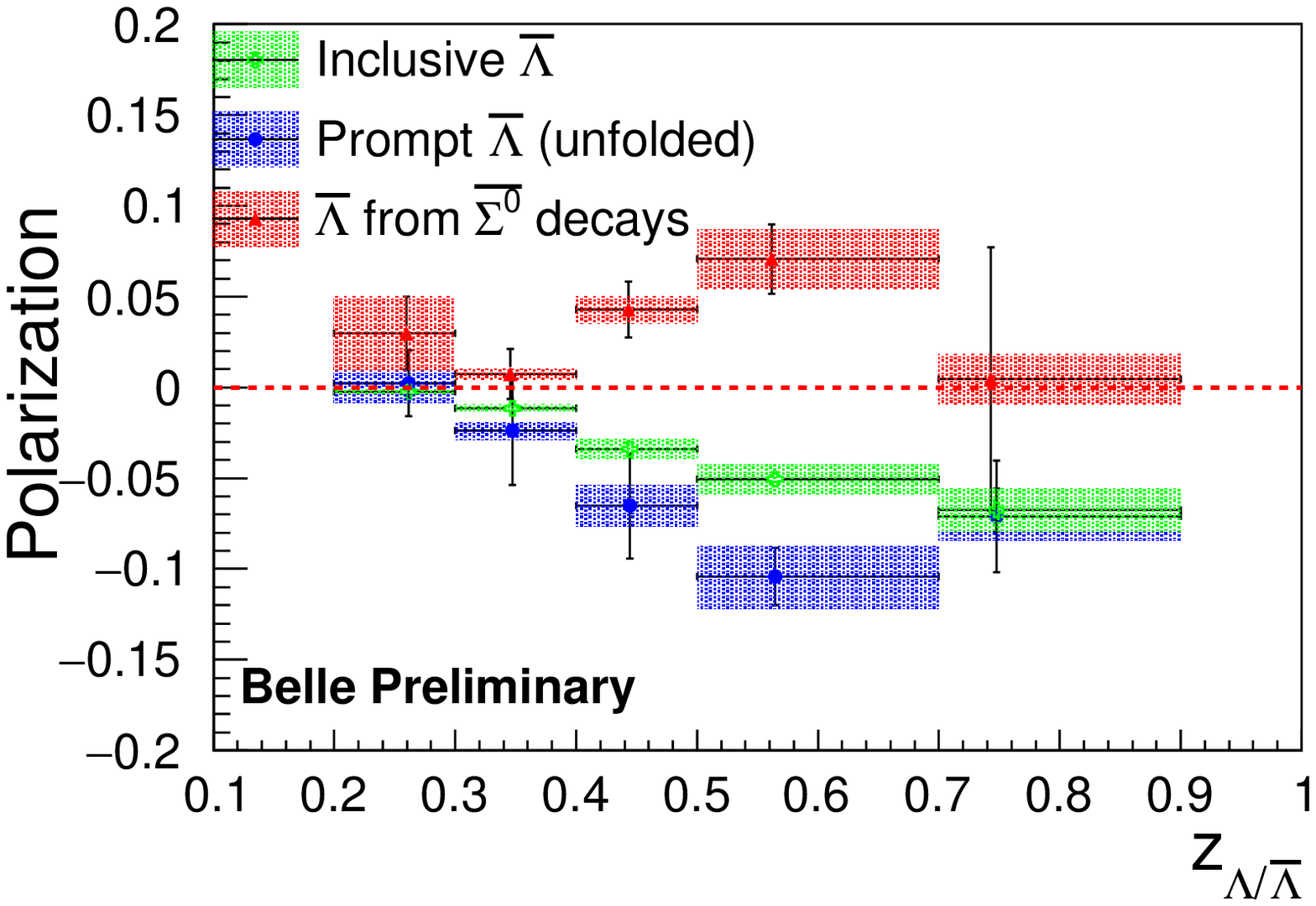}
\caption{The unfolded transverse polarizations of $\Lambda$ (blue dots) and $\Lambda$ from $\Sigma^0 \rightarrow \Lambda + \gamma$ decays (red triangles) and comparing to the original polarizations observed for inclusive $\Lambda$ as a function of $z_\Lambda$ in the thrust frame. The error bars show statistical uncertainties and the shaded areas show the systematic uncertainties. For the unfolding, the decay mode $\Lambda_c \rightarrow \Lambda + \pi^{+}$ is also considered.} 
\label{fig:pol_forSigma0_sys_lcmode0}
\end{center}
\end{figure}

The above results show the transverse polarization for inclusive $\Lambda$, including directly- and indirectly-produced $\Lambda$. In fact, in addition to the directly-produced $\Lambda$, this sample contains sizable contributions from feed-down, where the observed $\Lambda$ originates from the electroweak decay of a heavier resonance. (Note that we consider strong decays, such as the $\Sigma^*$, to be part of the directly-produced $\Lambda$ sample.) The dominant contributions come from $\Sigma^0$ and $\Lambda_c$, where we estimate the contributions to our sample to be on average $23\%$ and $20\%$, respectively.
Since there is no clear a priori knowledge of the polarization of the produced $\Sigma^0$ and $\Lambda_c$ or the daughter $\Lambda$, our approach is to use $\Sigma^0$- and $\Lambda_c$-enhanced samples to unfold the polarization of the direct contribution from the feed-down contributions.
The measured polarization is
\begin{equation}~\label{eq:bkgunfold}
P^{\rm mea} = (1 - \sum_{i}F_{i}) P^{\rm true} +  \sum_{i}F_{i}P_{i},
\end{equation}
where $P^{\rm true}$ is the polarization of directly-produced $\Lambda$, $F_i$ is the fraction of the $i^{th}$ source of $\Lambda$ and $P_{i}$ is the polarization of $\Lambda$ associated with the $i^{th}$ process. 
To construct a $\Sigma^0$ enhanced sample, we use the observed $\Lambda$ to reconstruct a $\Sigma^0$ in the channel $\Sigma^0 \rightarrow \Lambda + \gamma$, which practically saturates the branching ratio ($\mathcal{B}$) of the $\Sigma^0$~\cite{PDG}. The situation for the $\Lambda_c$ is more complicated, since there are several channels of comparable magnitude that decay with a $\Lambda$ in the final state. Most are three- or more- body decays, which we cannot hope to use for unfolding with a reasonable efficiency. Here, we use the channel $\Lambda_c \rightarrow \Lambda + \pi^+$ with $\mathcal{B} = 1.07 \pm 0.28 \%$~\cite{PDG}. 
Note that, therefore, the assumption in the following is that the $\Lambda$ polarization depends only weakly on the decay of the $\Lambda_c$.
The $\Sigma^0$- and $\Lambda_c$-enhanced data sets are analyzed using the procedure described above and, together with the original $\Lambda$ sample, used to do the unfolding based on Eq.~\ref{eq:bkgunfold}. The $F_{i}$ are obtained from MC. 
Unfortunately, due to the limited statistics, the unfolding is only done for five $z_{\Lambda}$ bins with boundaries at $z_{\Lambda}$=[0.2,  0.3, 0.4, 0.5, 0.7, 0.9] in the thrust frame and we cannot consider the transverse momentum dependence or any light hadron in the other hemisphere. The results are shown in Fig.~\ref{fig:pol_forSigma0_sys_lcmode0}. 
The transverse polarization of $\Lambda$ from $\Sigma^0$ decays is found to have the opposite sign to direct quark fragmentation. The magnitude of the amplitude is about half. This might be explained by the orbital angular momentum carried by the $\gamma$. 
While the $\Lambda_c$-enhanced sample enters in the unfolding with respect to the $\Lambda$ and $\Sigma^0$, the statistical uncertainties are too large to draw any conclusions about the magnitude of the $\Lambda_c$ feed-down itself. However, within the large uncertainties, it is consistent with zero.

Systematic uncertainties are estimated through various dedicated studies.  
To estimate the smearing or bias effects due to detector acceptance and resolution, we use a re-weighted sample to generate nonzero transverse polarization of $\Lambda$ in a simulation sample.
The input polarization is modeled as a function of $z$ and $p_{\rm t}$ to be near the observed asymmetries and varied to envelop the observed asymmetries. As expected, we find a significant smearing effect in the thrust frame due to the uncertainty of the real thrust axis. The correction factors for the magnitude of the measured transverse polarization vary from 1.0 to 1.3, depending on $z_\Lambda$. The uncertainties of the scale factors, which arise from the statistical limitation of the MC sample, are assigned as systematic errors. 
The correction factors in the hadron frame are consistent with 1.0, which is expected, since the direction of charged tracks is reconstructed very well.

We vary the shape used to describe the background contributions under the $\Lambda$ mass peak by changing the functional form fitted to the sidebands. Thus, the scale factor of sidebands is varied when the sideband subtraction is applied to $\cost$ distribution. The resulting difference is assigned as a systematic uncertainty.


While likelihoods of the fits in Eq.~\ref{eq:main}, as estimated from the fit $\chi^2$ values, are consistent with the fitted model, we estimate the systematics from possible non-linear $\cost$ contributions. For this purpose we added a second-order term of the form $1+p_{0}\cost + p_{1}{\rm{cos}}^{2}\theta$, where $p_{0}$ and $p_{1}$ are free parameters. The difference in the extracted polarizations from the nominal values is assigned as a systematic error.  

The uncertainties from the decay parameters $\alpha$ are assigned as systematic uncertainties as well. 
Several other studies are performed to check for possible bias in our measurements. Two important checks are explained as follows. First, the reference axis is replaced by  $\hat{n}' \propto \hat{p}_{\Lambda} \times \hat{n}$, which is still normal to $\Lambda$ direction but in the $\Lambda$ production plane; zero polarization is observed, as expected.  
Second, we use event mixing by reconstructing $\Lambda$ candidates using a proton and a pion from different events. The obtained polarizations are again consistent with zero. All systematic uncertainties are added in quadrature, and are shown using the shaded areas in Figs.~\ref{fig:pol_vszpt_sys}, ~\ref{fig:pol_zlzh_sys}, ~\ref{fig:pol_zlzh_sys_href} and ~\ref{fig:pol_forSigma0_sys_lcmode0}. 

In summary, we have studied the transverse polarization of $\Lambda (\bar{\Lambda})$ in the inclusive process $e^+e^-\to \Lambda (\bar{\Lambda}) + X$ and $e^+e^-\to \Lambda (\bar{\Lambda})  + \pi^{\pm}(K^{\pm}) + X$ with the data collected by Belle. A clear nonzero transverse polarization is observed, which is the first such observation in $e^+e^-$ annihilation. 
Its magnitude as a function of $z_\Lambda$ and $p_{\rm t}$ is presented. By selecting an identified light hadron in the opposite hemisphere, we obtain some sensitivity to the flavor dependence of the observed $\Lambda$ polarization. Clear contributions from different combinations of $\Lambda(\bar{\Lambda}) + h$ are seen.  
Furthermore, we attempt to separate the contributions for directly-produced $\Lambda$s and those from $\Sigma^0$  and $\Lambda_c$ decays.
The results presented in this report provide rich information about the transverse polarization of $\Lambda$ and will be very useful to understand the spin structure of hyperons.

\begin{acknowledgments}
We thank the KEKB group for the excellent operation of the
accelerator; the KEK cryogenics group for the efficient
operation of the solenoid; and the KEK computer group,
the National Institute of Informatics, and the 
PNNL/EMSL computing group for valuable computing
and SINET4 network support.  We acknowledge support from
the Ministry of Education, Culture, Sports, Science, and
Technology (MEXT) of Japan, the Japan Society for the 
Promotion of Science (JSPS), and the Tau-Lepton Physics 
Research Center of Nagoya University; 
the Australian Research Council and the Australian 
Department of Industry, Innovation, Science and Research;
Austrian Science Fund under Grant No.~P 22742-N16 and P 26794-N20;
the National Natural Science Foundation of China under Contracts 
No.~10575109, No.~10775142, No.~10875115, No.~11175187, and  No.~11475187; 
the Ministry of Education, Youth and Sports of the Czech
Republic under Contract No.~LG14034;
the Carl Zeiss Foundation, the Deutsche Forschungsgemeinschaft
and the VolkswagenStiftung;
the Department of Science and Technology of India; 
the Istituto Nazionale di Fisica Nucleare of Italy; 
National Research Foundation (NRF) of Korea Grants
No.~2011-0029457, No.~2012-0008143, No.~2012R1A1A2008330, 
No.~2013R1A1A3007772, No.~2014R1A2A2A01005286, No.~2014R1A2A2A01002734, 
No.~2014R1A1A2006456;
the Basic Research Lab program under NRF Grant No.~KRF-2011-0020333, 
No.~KRF-2011-0021196, Center for Korean J-PARC Users, No.~NRF-2013K1A3A7A06056592; 
the Brain Korea 21-Plus program and the Global Science Experimental Data 
Hub Center of the Korea Institute of Science and Technology Information;
the Polish Ministry of Science and Higher Education and 
the National Science Center;
the Ministry of Education and Science of the Russian Federation and
the Russian Foundation for Basic Research;
the Slovenian Research Agency;
the Basque Foundation for Science (IKERBASQUE) and 
the Euskal Herriko Unibertsitatea (UPV/EHU) under program UFI 11/55 (Spain);
the Swiss National Science Foundation; the National Science Council
and the Ministry of Education of Taiwan; and the U.S.\
Department of Energy and the National Science Foundation.
This work is supported by a Grant-in-Aid from MEXT for 
Science Research in a Priority Area (``New Development of 
Flavor Physics'') and from JSPS for Creative Scientific 
Research (``Evolution of Tau-lepton Physics'').
\end{acknowledgments}

\bibliography{lambda}
\end{document}